\documentclass[a4paper,final,onecolumn,10pt]{IEEEtran}
\usepackage{graphicx}
\usepackage{bm}
\usepackage[USenglish]{babel}
\usepackage{amsmath}
\begin{document}
\parindent=0mm
\title{Cyclic Correlation of Diffuse Reflected Signal with Glucose Concentration and scatterer size} 
\author{Jitendra Solanki, Pratima Sen$^{\dagger}$, Joseph Thomas Andrews and Kamal Kishore Thareja$^{\ddagger}$}
\maketitle
\begin{center}
\vspace{-1cm}
Department of Applied Physics, Shri G S Institute of Technology \& Science,Indore - 452 003 India.\\ 
$^{\dagger}$Laser Bhawan, School of Physics, Devi Ahilya University, Khandwa Road, Indore - 452 007 India.\\
$^{\ddagger}$KK Pathology Lab., Chanakya Complex, Malwa Mill Chourah, Indore - 452 003 India.\\
Corresponding author email: jtandrews@sgsits.ac.in
\vspace{1cm}
\end{center}

\begin{abstract} 
The utility of optical coherence tomography signal intensity for measurement
of glucose concentration has been analysed in tissue phantom and blood samples
from human subjects. The diffusion equation based calculations as well as
in-vivo OCT signal measurements confirms the cyclic correlation of signal intensity 
with glucose concentration and scatterer size.

\end{abstract}
\section{Introduction}
Everyday, almost 150 million people world wide face the problem of diabetic metabolic control. Both the hypo- and hyper-glycemic conditions of patients have fatal consequences and warrant blood glucose monitoring at regular interval. Existing monitors of blood glucose can be widely classified  into three classes
viz., invasive, minimally invasive, and noninvasive. Invasive monitoring require small volume of blood and they are inappropriate for continuous monitoring of blood glucose. Minimally invasive monitors analyze tissue fluid, and skin injury is minimally. On the other hand,  noninvasive devices are painless and skin injury is absent. 

Many research groups are working on to develop a real time noninvasive tool for monitoring blood glucose at clinical level. Recent review by Bazaev and Selishchev \cite{1} discuss various noninvasive methods for blood glucose measurement and monitoring. Existing non-invasive  blood glucose monitoring work on various modalities such as absorption spectroscopy \cite{2},  optical activity and polarimetry \cite{3,4}, Optical Coherence Tomography (OCT) \cite{5,6},  bioimpedance spectroscopy \cite{2}, fluorescence \cite{7}, etc. The other optical techniques include Raman spectroscopy \cite{8} and reflectance spectroscopy \cite{9}. However, due to optical interference, poor signal strength, and calibration issues, optical methods still face many challenges \cite{10}. The measurement of blood glucose level using OCT technique exhibits large fluctuations due to motion artifacts or other physiological and environmental conditions.
\cite{5,6}. 

Optical coherence tomography is a nondestructive technique that examines the internal structure of superficial layers of biological tissues. It is based on interferometric recording of near-infrared light backscattered from the point of study, which could carry information. Conventionally, in OCT backscattered light is collected, measured, and integrated to assemble images \cite{11}. 

The impetus for the present paper comes from two reasons. Firstly, to correlate the glucose concentration with experimentally observed diffuse reflectance signal from OCT. Secondly to examine the usability of OCT signal intensity for measurement of glucose concentration. Accordingly, the paper has been divided into three parts. Section two deals with the theoretical analysis where Mie scattering theory and diffusion equation are used  to understand the dependence of solute concentration on OCT signal intensity in a turbid medium. In the  third section, experimental results are exhibited for various samples of aqueous solutions of glucose dissolved in tissue phantom as the scatterers. The experiment was also performed with blood samples from voluntary blood donors. In the fourth section, the numerical calculations were carried out for various glucose concentrations in aqueous solution having different particle sizes as scatterers using the theoretical analysis. It is observed that a cyclic correlation of diffuse reflected signal exist as a function of the glucose concentration and size of the scatterers. To support the theoretical results, experimental observations are also presented.

\section{Theoretical Formulations} 
Blood is a turbid medium consisting of extracellular fluid (ECF) and  various
scatterers e.g. red blood cells (RBC), white blood cell (WBC), etc. Multiple scattering from RBC gives rise to diffusion of light from blood.
The average diffuse intensity ($U_d$) is related to the scattering cross section via the relation 
\cite{6}
\begin{equation}
U_d(r)=U_d(0)\exp(-k_d r). 
\end{equation}
$k_d$ is the spatial decay constant and is given by $k_d$~(~=~$\rho \sqrt{3\sigma_a \sigma _{tr}}$~) where, $\rho$ is the number density of scatterers in the medium, $\sigma_{tr} $ is the transport cross section and is given by $\sigma_{tr}\,[=\sigma_s(1-g)+\sigma_a]$ with $\sigma_s$ being the scattering coefficient, $g$ is the anisotropy parameter and $\sigma_a$ is the absorption coefficient.

The scattering cross sections can be calculated from the Mie scattering theory as \cite{12}
\begin{equation}
\sigma_s=\frac{2a^{2}\pi \left|\sum_{n}{(2n+1)(-1)^{n}(a_n-b_n)}\right|^{2}}{\alpha^{2}},
\end{equation}
where, 
\[a_n=\frac{\psi_n(\alpha)\psi_n^{'}(\beta)-m\psi_n(\beta)\psi_n^{'}(\alpha)}{\xi_n(\alpha)\psi_n^{'}(\beta)-m\psi_n(\beta)\xi_n^{'}(\alpha)}\]
and 
\[b_n=\frac{m\psi_n(\alpha)\psi_n^{'}(\beta)-\psi_n(\beta)\psi_n^{'}(\alpha)}{m\xi_n(\alpha)\psi_n^{'}(\beta)-\psi_n(\beta)\xi_n^{'}(\alpha)}\]
with $\alpha=k_da$, $k_d$ being the propagation constant of the radiation in ECF and $a$ is the scatterer size. 
$\beta\,(=mk_da)$ with $m$ being the relative refractive index of the scatterer (RBC in the present case). Also,  $\psi_n(\alpha)=\alpha j_n(\alpha)$ and $\xi_n(\alpha)=\alpha h^{'}_n(\alpha)$, where $j_n(\alpha)$ and $h_n(\alpha)$ represent the spherical Bessel and Hankel functions, respectively.
With the knowledge of these parameters we can calculate the path length resolved diffuse reflectance as  
\begin{equation}
R(L_s)=\frac{1}{U_d(0)}\frac{\partial }{\partial L} U_d(L)|_{L=L_s}  \label{dif},
\end{equation}
where, $L_s(=L_{s0}+\sum _{i=0}^{\infty}L_{si})$ is the round trip path length to the sample surface and total path length within the sample that accumulates during each scattering event \cite{6}. A theoretical estimate of the OCT signal intensity can be made from the diffuse reflectance as follows: In an OCT set up the interference between the light scattered from the sample arm and reference arm is detected. The light intensity at the detector is given by 
\begin{equation}
I_d(\tau)=\left\langle \left|\int_{L_s}^{\infty}E_{s}^{,}(t, L_s)dL_s+E_r(t,\tau)\right|^{2}\right\rangle,
\label{id}\end{equation}
where, $\tau[=\Delta L/c]$ is the time delay corresponding to the round-trip optical path length between two beams and $E_{s}^{,}(t, L_s)$ is the path length resolved field amplitude. $E_r$ is field amplitude from the reference mirror. 

The OCT signal obtainable from a turbid medium is a convolution of path length resolved diffuse reflectance and the low coherence function arising from the coherence property of the laser source. Accordingly, the OCT signal may be rewritten as, 
\begin{equation}
I_d(L_s)=I_r [R(L_s)^{1/2}\otimes C(L_s)],
\end{equation} 
where, $I_d(L_s)$ is the OCT signal detected by a photodiode, $I_r$ is the signals from reference arm. $C(L_s)$ is low-coherence function of the source. 

The change in the intensity of OCT signal  vary in accordance with the square root of the variations in the diffuse reflectance. A close observation of equations (1) to (5) clearly shows that the diffuse reflectance $R(L_s)$ is intimately related to the scattering cross section $\sigma_s$, via equation (2). Also equation (2) suggests that the scattering cross section is a function of (i) scatterer size and (ii) relative refractive Indies of the scatterer and the fluid in which the scatterer is dispersed. 
Consequently, one can expect a variation in the OCT signal intensity with the change in the refractive index of ECF as well as change in the size of RBC.

The amplitude of the OCT signal $A(L_s)$ is proportional to  the diffuse reflectance  ($R(L_s)$) and is given by 
\begin{equation}
A(L_s)=\frac{\Delta U_d(r)}{U_d(r)}|_{L=L_s}.\label{amp}
\end{equation}
It is worth noticing that the RBC cells, which act as source of scatterer in blood have various sizes in the  subjects under study. Consequently, the utility of OCT setup for examining glucose concentration in blood requires
the knowledge of weighted average of the signal ($AW(L_s)$). Under the Gaussian
distribution approximation, it is given by \cite{1926}
 
\begin{equation}
AW(L_s)=\frac{1}{n}\sum_n\frac{\Delta U_d(r)}{U_d(r)}\exp\left[-\left(\frac{S_n-S_0}{\Delta S}\right)^2\right],\label{wamp}
\end{equation}
with $S_0$, $S_n$ and $\Delta S$ as the size of RBC, mean size of RBC and
variance in size, respectively. The reported  values of RBC in healthy subjects
are $S_0\approx7.5 \mu$m and  $\Delta S\approx1 \mu$m \cite{19}. 

\section{Experiment} 
\begin{figure}[htb]
\begin{center}
\includegraphics[width=.9\columnwidth]{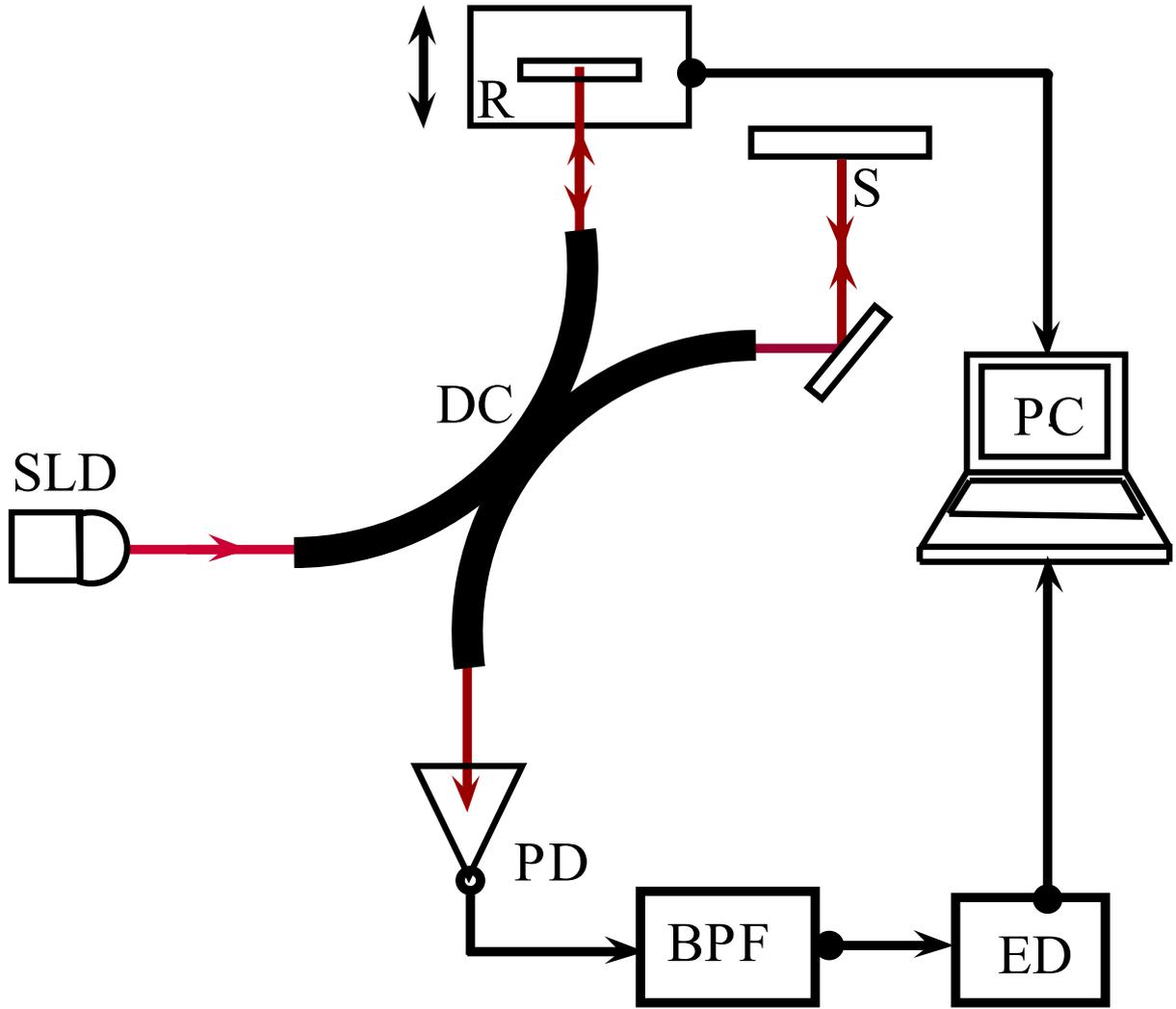}
\caption{Schematic of the  Optical Coherence Tomography setup. SLD~-~Superluminescent
diode, DC~-~3dB bi-Directional Coupler, PD-Photo Diode, S~-~ Sample, R-Reference arm with scanning assembly,
BPF-Band pass filter, ED-Envelope detector.}
\end{center}
\end{figure}
The optical coherence tomography setup used in the present report is developed in house, using a fiber optic  Michelson interferometer. We use  superluminescent diode (SLD, Superlum SLD-371) as the light source, which
is a  broad band source with a center wavelength of 841 nm and line width
of 52nm.  The selection
of wavelength in the near infrared region has the advantages of
large scattering   coefficient and low absorption coefficient in tissues. However, these coefficients are altered when index mismatch occurs between ECF and cells.
 
\subsection{Sample Preparation} The primary goal of the present work is to find the correlation between the amplitude of OCT signal with concentration of glucose. A systematic measurements were carried out with the following
samples;
 (i)~tissue phantom based as intralipid \cite{13} and (ii) blood samples from voluntary human subjects. 

\subsection{Tissue Phantom}  
 Intralipid is an emulsion of soy bean oil, egg phospholipids and glycerin \cite{14}. Intralipid is widely used in optical experiments to simulate the scattering properties of biological tissues and 
can be use as a good scatterer as RBC in blood. The major advantages of intralipid are its well known optical properties and the similarity of its microparticles to lipid cell membranes and organelles that constitute the source of scattering in biological tissue \cite{15}. 
We use intralipid  as a tissue phantom that provides the backscattered component. Average size of scatterers in intralipid  measured using confocal microscope is found to be 3.5 $\mu$m.

Using standard approach, we generate glucose concentrations of 0~--~200 mg/dl in 1ml sample tissue phantom for experimental observations.  This range covers hypoglycemic ($<$ 80 mg/dl), normal (80-120 mg/dl) as well as hyperglycemic ($>$120 mg/dl) conditions. A settlement time of 2~minutes was given after addition of glucose to the intralipid solution so that the added glucose could alter the scattering properties of the medium. Simultaneously, the change in refractive index of the solution is also monitored using Abbe's
refractometer.  Measurement was done using fixed volume 1ml of tissue phantom and diluted to 0.01\% in distilled water (100ml) and concentrated glucose solution prepared by dissolving 200 mg glucose in 100 ml distilled water. The concentration of glucose in intralipid  is increased in steps of  10mg/dl
using U-TEK Chromatography syringe with least count of 5$\mu$l. 

\subsection{Blood samples} 
Blood samples are  collected from Pathological Laboratory. In order to optimize the setup with human subjects, blood samples were collected from various
voluntary subjects. Conventional pathological method known as GOD/POD (Glucose Oxydised / Peroxydised) using semi-automated blood analyser was used simultaneously to measure the exact value of blood glucose. 
A smaller part (few$\mu l$) of the glucose solution was used in the OCT setup. 
Small quantity of 4 $\mu$l  blood taken using a Eppendorf Pipette. 

\section{Results}
\begin{figure}[htb]   
\begin{center}
\includegraphics[width=0.9\columnwidth]{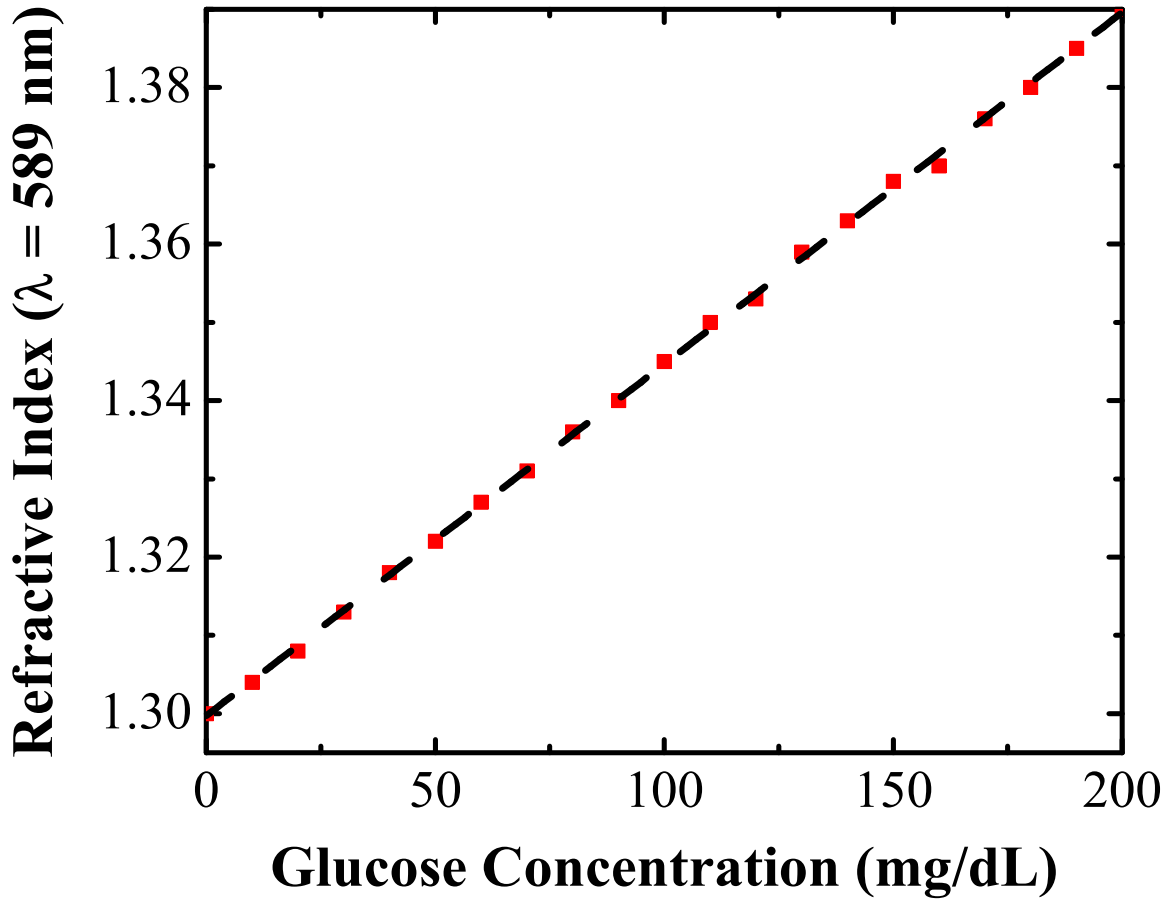}
\caption{Variation of refractive index of intralipid with Glucose concentration
measured using Abbe's refractometer.}\label{Abbe}
\end{center}
\end{figure}

To confirm the changes in the optical properties of the solution with increasing glucose concentration in intralipid, Abbe refractometer was used to obtain the refractive index of tissue phantom and found to be 1.407. As discussed earlier,  addition of 50$\mu l$ glucose to  intralipid solution of volume 1ml increases the glucose concentration by  10mg/dl. At the same time, it  reduces the index matching between the scatterer and the medium. The measurements of refractive index and scattering coefficients
using Abbe refractometer and OCT, respectively are made simultaneously
to avoid the temporal fluctuations in values. Results obtained with refractometer are displayed in Fig. \ref{Abbe}. A linear relation is obtained between the glucose concentration and the refractive index. The data from Fig. \ref{Abbe} is used to obtain an empirical relation between the glucose concentration  and the refractive index.
 
\begin{figure}[htb]   
\begin{center}
\includegraphics[width=0.9\columnwidth]{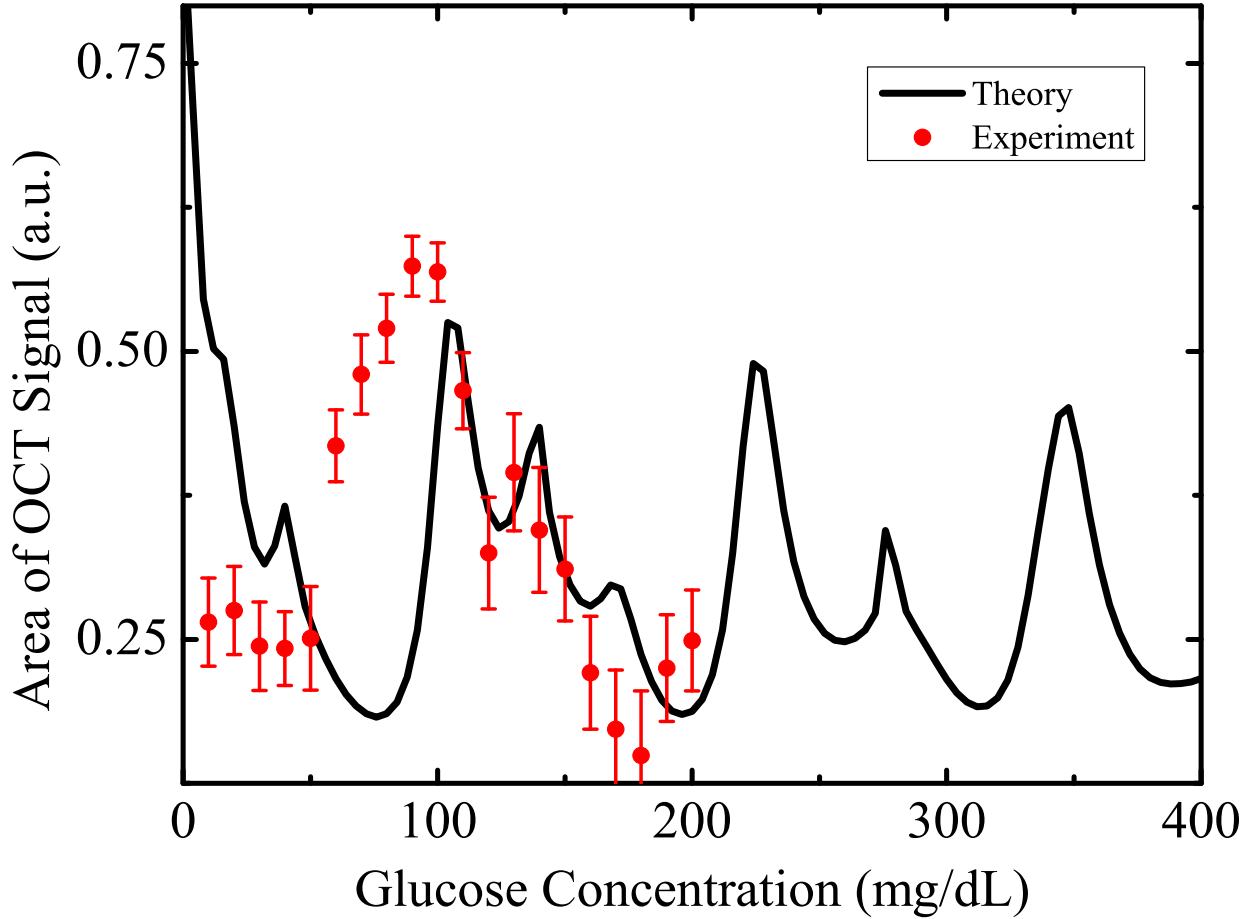}
\caption{OCT signal amplitude obtained in a sample of tissue phantom with different glucose concentrations are shown as filled circles with error bars. The solid line is the weighted average obtained theoretically using equation (\ref{wamp}).}\label{Tispha}
\end{center}
\end{figure}

OCT signal amplitudes  obtained with  tissue phantom and for various glucose concentrations (10-200mg/dl) are shown in Fig. \ref{Tispha}. In order to reduce the noise in OCT signal, we displayed the normalised average area of scattered signal within the coherence length and shown as dots.  The OCT signal amplitude as obtained now is numerically equal to the average diffuse intensity $I_d$ defined in eq. (\ref{id}). The theoretical fit shown as solid line in the  Fig. \ref{Tispha} is the weighted average as discussed earlier. The results
indicate an oscillatory nature of backscattered component of the OCT signal with glucose concentration.  These observations with tissue phantom show
a cyclic correlation between glucose concentration and OCT signal amplitude
 and are in good agreement with the theoretically calculated values of diffuse reflectance
(solid line). Both the curves exhibit similar nature which confirms the usability of present theoretical calculations to correlate them with blood glucose levels in human subjects. In order to confirm the same with clinical measurements with human subjects, we collected samples of voluntary donors from pathological laboratory and the results are discussed below.   

\begin{figure}[htb]   
\begin{center}
\includegraphics[width=0.9\columnwidth]{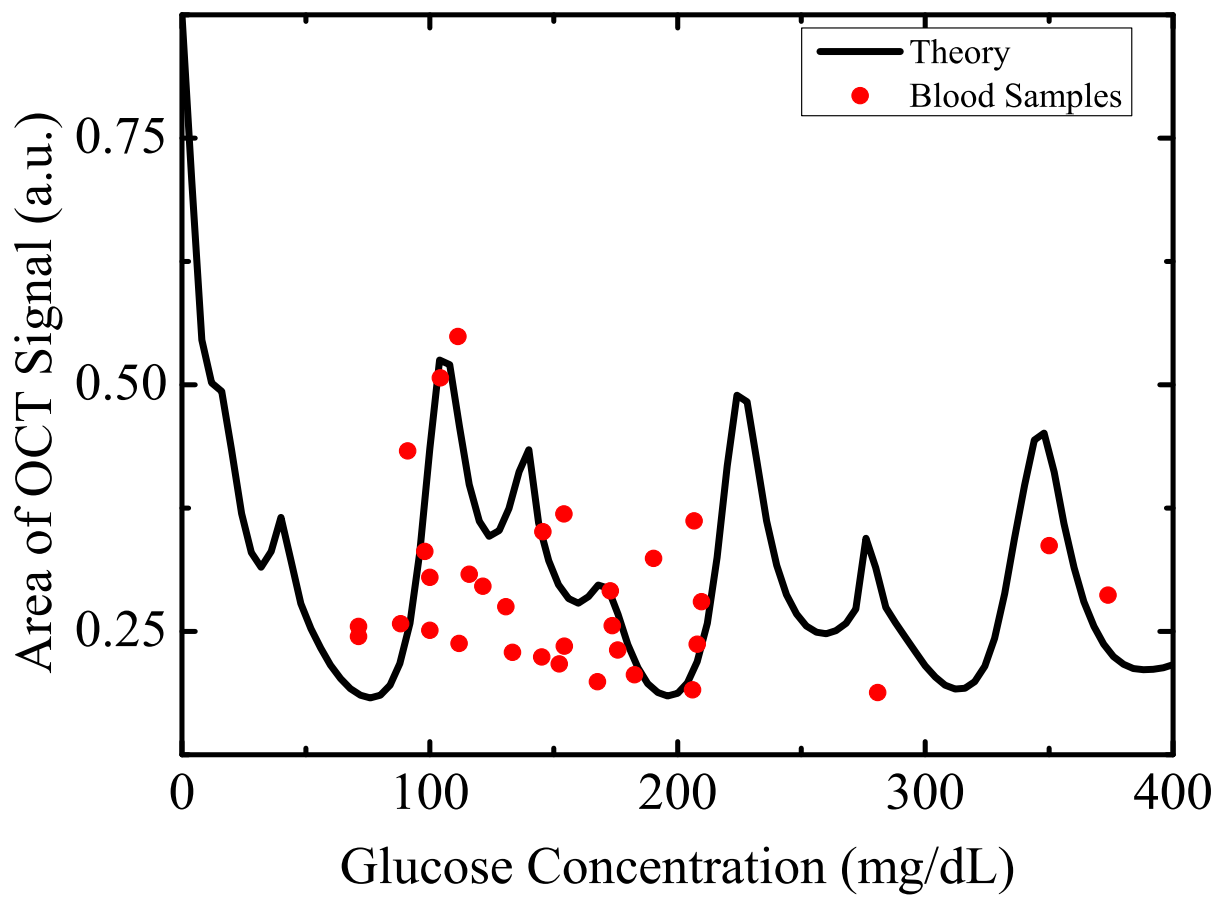}
\caption{The filled circles represent the amplitude of OCT signal obtained from various blood samples of voluntary donors having different glycemic levels while the solid curve is the weighted average obtained theoretically using equation~(\ref{wamp}).}\label{blood}
\end{center}
\end{figure}

Simultaneous measurements of blood glucose was carried out using chemical methods as described in previous sections. The correlation between blood glucose level of human subjects and the amplitude of the OCT signal are represented in Fig.~\ref{blood}. The curve definitely shows a cyclic correlation, similar to the results predicted from eq. (\ref{id}) and Fig.~\ref{Tispha}. 

We therefore, find that the cyclic correlation of glucose concentration
with OCT signal restricts the OCT signal intensity based glucose measurements in human subjects. 

To conclude, although it is well known that the optical properties of tissue vary with  glucose concentration and the index mismatch between the ECF and scatterers. 
There are many factors which will affect the OCT\ signal intensity. For
example  the refractive index of ECF will change with change in any types of soluble impurities (example : urea) and the change in the scatterer size. Accordingly, well defined predictions for glucose concentration measurements
using OCT signal intensity are not possible  because of the non-uniform size distribution of various particles as well as presence of creatinine, lactic acid, serum albumin, and NaCl, Urea, etc. in blood. Therefore, the results  presented here may not be used as a tool for estimation of blood glucose levels, however, it throws light on the understanding of light scattering from blood samples, in the presence of various types of scatterers differing in physical and optical properties.

\section*{Acknowledgments} 
The authors thank financial support received  from UGC, New Delhi. The authors also thank Prof. P. K. Sen, SGSITS for fruitful discussions. JS acknowledges the support received from  Prof. S. Kumbhaj, SGSITS and Prof. A. Mishra, DAVV Indore.


\begin{thebibliography}{99}

\bibitem{1} N. A. Bazaev and S. V. Selishchev, Noninvasive Methods for Blood Glucose Measurement, Biomed, Eng, {2007}, 41, 40-48.
\bibitem{2} C.F. Amaral, M. Brischwein, B. Wolf, Multiparameter techniques for noninvasive measurement of blood glucose, Sensors and Actuators B, {2009}, 140, 12-16.

\bibitem{3}B. Rabinovitch, W. F. March, R. L. Adams, Noninvasive glucose monitoring of the aqueous humor of the eye, Part 1, Measurement of very small optical rotations, Diabetes Care, {1982}, 5, 254-258.
\bibitem{4} G. L. Cote, M. D. Fox, R. B. Northrop: Noninvasive optical polarimetric glucose sensing using a true phase measurement technique, IEEE, Trans, Biomed, Eng, {1992}, 39, 752-756.

\bibitem{5} K. V. Larin, M. Motamedi, M. S. Eledrisi, R. O. Esenaliev, Noninvasive Blood Glucose Monitoring With Optical Coherence Tomography, Diabetes Care, {2002},25, 2263-2267.
\bibitem{6} R. Poddar, S. R. Sharma, J. T. Andrews and P. Sen, Study of correlation between glucose concentration and reduced scattering coefficients in turbid media using optical coherence tomography, Current Science, {2008}, 95, 2.

\bibitem{7} J. C. Pickup, F. Hussain, N. D. Evans, O. J. Rolinski, D. J. S. Birch, Fluorescence-based glucose sensors, Biosensors and Bioelectronics, {2005}, 20, 2555-2565.
\bibitem{8} J. R. McNichols and L. G. Cote, Optical glucose sensing in biological fluids: An Overview, Journal of Biomedical Optics , {2000}, 5, 5-16.

\bibitem{9} S. F. Malin, T. L. Ruchiti, T. B. Blank, S. U. Thennadil, and S. L. Monfre,, Noninvasive prediction of glucose glucose by near infrared diffuse reflectance spectroscopy, Clini. Chem., {1999}, 45, 1651-8.
\bibitem{10} V. Ashok, A. Nirmalkumar, and N. Jeyashanthi, A Novel Method for Blood Glucose Measurement by Noninvasive Technique Using Laser, International Journal of Biological and Life Sciences, {2010}, 6, 3.
\bibitem{11} V. V. Sapozhnikova, D. Prough, R. V.Kuranov, Influence of Osmolytes on in vivo Glucose Monitoring using OCT, {2006},231, 1323-1332.
\bibitem{12} A. Ishimaru, Wave Propagation and scattering in Random Media, Academic Press, New York, {1978}.
\bibitem{1926} D. N. medearis and G. E. Minot, Studies on red blood cell
diameter, J. Clin. Invest., 1926, 229, 541-556.
\bibitem{19} S. N. Thennadill, J. L. Rennert, B. J. Wenzel, K. H. Hazen, T. L. Ruchti, M. B. Block, Construction of glucose concentration in interstitial fluid, and capillary and venous blood during rapid changes in blood glucose levels. Diabetes Technol Ther. {2001}, 3, 357-65.  
\bibitem{13} R. Srinivasan and M. Singh, Laser Backscattering and Transillumination Imaging of Human Tissues and Their Equivalent Phantoms, IEEE Trans. on Biomed. Engin., {2003}, 50,6.
\bibitem{14} H. J. V. Staveren, C. J. M. Moes, J. V. Marle, S. A. Prahl, and M. J. C. V. Gemert, Light scattering in Intralipid-10 percent in the wavelength range of 400-1100 nm, Applied Optics, {1991}, 30, 31.
\bibitem{15} B. F. Kennedy, S. Loitsch, R. A. McLaughlin, L. Scolaro, Paul Rigby, and D. D. Sampson, Fibrin Phantom for use in optical coherence tomography, Journal of biomedical optics, {2010}, 15 , 1083-1088.
\end{thebibliography}
\end{document}